\documentclass[a4paper,10pt,twocolumn]{article} 
\usepackage{graphicx}
\usepackage{amsmath}
\usepackage{amssymb}
\usepackage{wrapfig}
\usepackage{color}
\usepackage[english]{babel}
\usepackage{enumerate} 
\usepackage{multirow} 
\usepackage[pdftex,
            pdfauthor={Yves-Henri Sanejouand},
            pdftitle={A pair of possible supernovae Refsdal in the Pantheon+ sample},
            pdfsubject={Strong lensing},
            pdfkeywords={Strong lens; Supernovae; White dwarf},
            pdfproducer={Latex with hyperref},
            pdfcreator={Texmaker}]{}
\usepackage{indentfirst}
\usepackage{hyperref} 
\begin{document}

\title{\textbf{A pair of possible supernovae Refsdal\\ in the Pantheon+ sample}}

\author{Yves-Henri Sanejouand\footnote{yves-henri.sanejouand@univ-nantes.fr}\\
        Facult\'e des Sciences et des Techniques, Nantes, France.} 
\date{September 7$^{nd}$, 2024}
\maketitle

\section*{Abstract}

On December 1980, supernova 1980N was discovered
in NGC 1316, a galaxy of the Fornax cluster.
Three months later, supernova 1981D was observed in the same galaxy.
The light curves of these two supernovae Ia were found
to be virtually identical, suggesting that they are 
images of the same event, the delay between them being
due to strong gravitational lensing.
If so, as anticipated by Sjur Refsdal, the distance 
to the lens can be determined accurately, 
namely, 90 $\pm$ 1 kpc,
meaning that it belongs to the outer halo of the Milky Way.

Interestingly, there is another pair of possible images in the Pantheon+ sample, namely, supernovae 2013aa and 2017cbv,
the distance to the lens being 
702 $\pm$ 1 kpc, that is, nearly the same as 
the distance to the Andromeda galaxy. 

In both cases, given the relatively large angle of deviation of the supernova light
by the lens, namely, 271" and 325", respectively, the lens has to be a compact object, with a mass to radius ratio over
150~$M_\odot R_\odot^{-1}$. It is likely to be an ultra massive white dwarf.  

\vspace{0.5cm}
\noindent
Keywords: Supernovae; Gravitational lensing; White dwarf; Local Group.

\section*{Introduction}

In 1979, 
a possible pair of images of a quasar
formed by a gravitational lens was described \cite{Weymann:79}.
Comparison of the light curves of these two images later
showed that there is a delay between them, 
of more than 400 days \cite{Schramm:96,Oscoz:01}.
As previously realized by Sjur Refsdal,
when such a delay is known, the mass of the lens
can be estimated while,
when the mass of the lens is known,
the Hubble constant can be determined \cite{Refsdal:64},
as well as other cosmological parameters \cite{Auger:09}. Based on these ideas,
the H0LiCOW project has for instance provided a value of the Hubble constant at the 2.4\% level \cite{Meylan:20},
in agreement with recent measurements of the SH0ES team \cite{Riess:22}.

In 2014, four images of a supernova, coined supernova Refsdal, were observed, the lens being an early-type galaxy in the cluster MACS J1149.5+2223 \cite{Tucker:15}.
A fifth image was observed with a delay of roughly one year,
the lens being this time the whole cluster \cite{Zitrin:16}.
Of course, delays can be longer than that.
For instance, in the case of supernova requiem,
a fourth image is expected to be observed in 2037, that is, 
twenty years after the first one \cite{Whitaker:21}.

In the later case, the three first images of the supernova
were found in archival Hubble Space Telescope imaging.
In the present study,
taking advantage of the recent increase in size of the 
most complete and best characterized set of supernovae Ia,
namely, the Pantheon+ sample \cite{Scolnic:22}, 
other such cases were looked for.   

\section*{Data}

Like in a previous study \cite{Sanejouand:23},
Pantheon+ data were 
retrieved from the PantheonPlusSh0es page of the github webserver\footnote{\url{https://github.com/PantheonPlusSH0ES/DataRelease}},
equatorial coordinates
being taken in the all\_redshifts\_PVs.csv file. 
The 798 supernovae Ia  
with a rather low redshift accuracy, namely,
with a redshift error of more than 0.0002,
were disregarded,
redshifts and corresponding errors
coming from the Pantheon+SH0ES.dat file.
For pairs of supernovae with same heliocentric redshifts, the SALT2 light curve stretch (x1) was also picked in the Pantheon+SH0ES.dat file.
When several stretch values were found for a given supernova\footnote{The magnitude of 127 supernovae of the Pantheon+ sample was measured up to four times.}, their average was considered.
Photometry, maximum light curve time (SEARCH\_PEAKMJD) and host mass (HOSTGAL\_LOGMASS) of each supernova 
were taken in the Pantheon+\_Data/1\_DATA/photometry directory.

\begin{figure}[t]
\vskip  0.05 cm
\includegraphics[width=7.8cm]{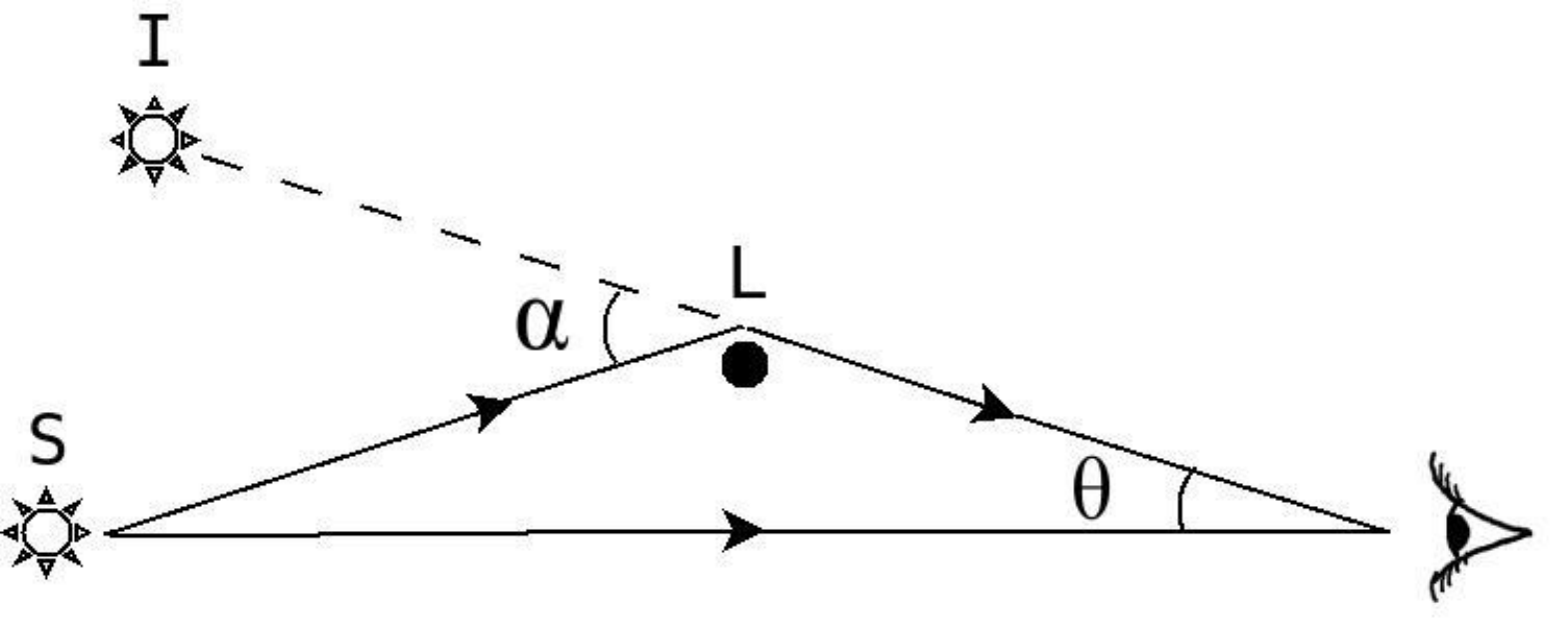}
\vskip -0.3 cm
\caption[]{The small-and-spherical lens approximation.
When the lens (L) is spherical, and the light deviation angle ($\alpha$) by the lens large enough, 
the light coming from a source (S) reaches the observer following two paths, an approximately straight one and a delayed one.
As a result, two images are seen by an observer, one (I) after the other (S),
separated by an angle $\theta$.
}
\label{Fig:small_lens}
\end{figure}

\section*{Distance to the lens}

Under the assumption that light arriving first from the source is not significantly perturbed by the gravitational field of the lens, with
\begin{equation}
\Delta_d = d_l + d_{ls} - d_s = c_0 \Delta t
\label{eq:delta}
\end{equation}
$\Delta t$ being the delay between the observations of the source and its image, 
$d_s$, the distance to the source, $d_l$, the distance to the lens, $d_{ls}$, the distance between the source and the lens  (see Figure \ref{Fig:small_lens}) and $c_0$, the speed of light,
$$
d_{ls}^2 = d_s^2 + d_l^2 - 2 d_s d_l \cos \theta 
$$
where $\theta$ is the angle between the source and its image, yields:
$$
d_l = \frac{d_s \Delta_d + \frac{1}{2} \Delta_d^2}{d_s (1 - \cos \theta) + \Delta_d}
$$
and since, in the present study, $\Delta_d \ll d_s$, with $\theta$ being small: 
\begin{equation}
d_l \approx \frac{2 \Delta_d}{\theta^2 + \frac{2 \Delta_d}{d_s}}
\label{eq:dapprox}
\end{equation}

\section*{Mass of the lens}

The angle of deviation of light by the lens, $\alpha$, is so that (Fig. \ref{Fig:small_lens}):
$$
\alpha \approx \theta \left( 1 + \frac{d_l}{d_{ls}} \right)
$$
that is, with eqn \ref{eq:delta}, and since $\Delta_d \ll d_l$:
\begin{equation}
\alpha \approx \theta \frac{d_s}{d_s - d_l}
\label{eq:alpha}
\end{equation}
On the other hand:
$$
\alpha = \frac{4 G M_l}{c_0^2 b}
$$
where $b$ is the distance of closest approach of the light to the lens, $M_l$ being the mass of the lens and $G$, the gravitation constant. Thus, with eqn \ref{eq:alpha}:
$$
\frac{M_l}{b} = \frac{c_0^2}{4 G} \frac{d_s}{d_s - d_l}
\theta 
$$
Though $b$ is not expected to be known, it is larger than $R_l$, the radius of the lens. So:
\begin{equation}
\frac{M_l}{R_l} > \frac{c_0^2}{4 G} \frac{d_s}{d_s - d_l}
\theta 
\label{eq:mr}
\end{equation}   

\begin{table*}[t]
\centering
\caption{Supernovae Refsdal of the Pantheon+ sample.}
\label{Table:refsdal} 
\hskip -0.1 cm
\begin{tabular}{|c|c|c|c|c|c|c|c|}
\hline
\multicolumn{6}{|c|}{Supernova} & \multicolumn{2}{|c|}{Lens} \\
\hline
\multirow{2}{*}{Host} & \multirow{2}{*}{z} & \multirow{2}{*}{first image} & \multirow{2}{*}{second one} & \multirow{2}{*}{$\theta$} & delay$^a$ & distance &  \multirow{2}{*}{$\frac{M_l}{R_l}$}$^b$ \\
 &  &  &  &  & (days) & (kpc) & \\ 
\hline 
 NGC 1316 & 0.005871 & 1980N & 1981D & 271" & 96 & 90~$\pm$~1 & 154 \\
 NGC 5643 & 0.004113 & 2013aa & 2017cbv & 325" & 1490 & 702~$\pm$~1 & 193 \\ 
\hline
\end{tabular}

$^a\pm 1$ day. $^b$Lower bound (solar units).
\end{table*}

\section*{Results}

\subsection*{Selection of candidates}
644 pairs of supernovae with same heliocentric redshift
were found in
the Pantheon+ sample,
that is, pairs of supernovae with redshifts differing by
no more than the sum of their respective errors.

Among these 644 pairs, only 12 have SALT2 stretches (x1)
differing by less than 0.02. 
For seven pairs, host masses are known for both of them.
However, for five of them, their host masses differ by more than the sum of their respective errors, host masses being compared on the logarithmic scale.

Thus, only two pairs of supernovae from the Pantheon+
sample have obviously same redshift, same SALT2 stretch 
and same host mass, namely, 1980N and 1981D, 2013aa and 2017cbv. Actually, both pairs are siblings \cite{Scolnic:20},
that is, they were observed in the same galaxy, namely,
NGC 1316 and NGC 5643, respectively.  

\begin{figure}[t]
\vskip -0.5 cm
\includegraphics[width=8.0 cm]{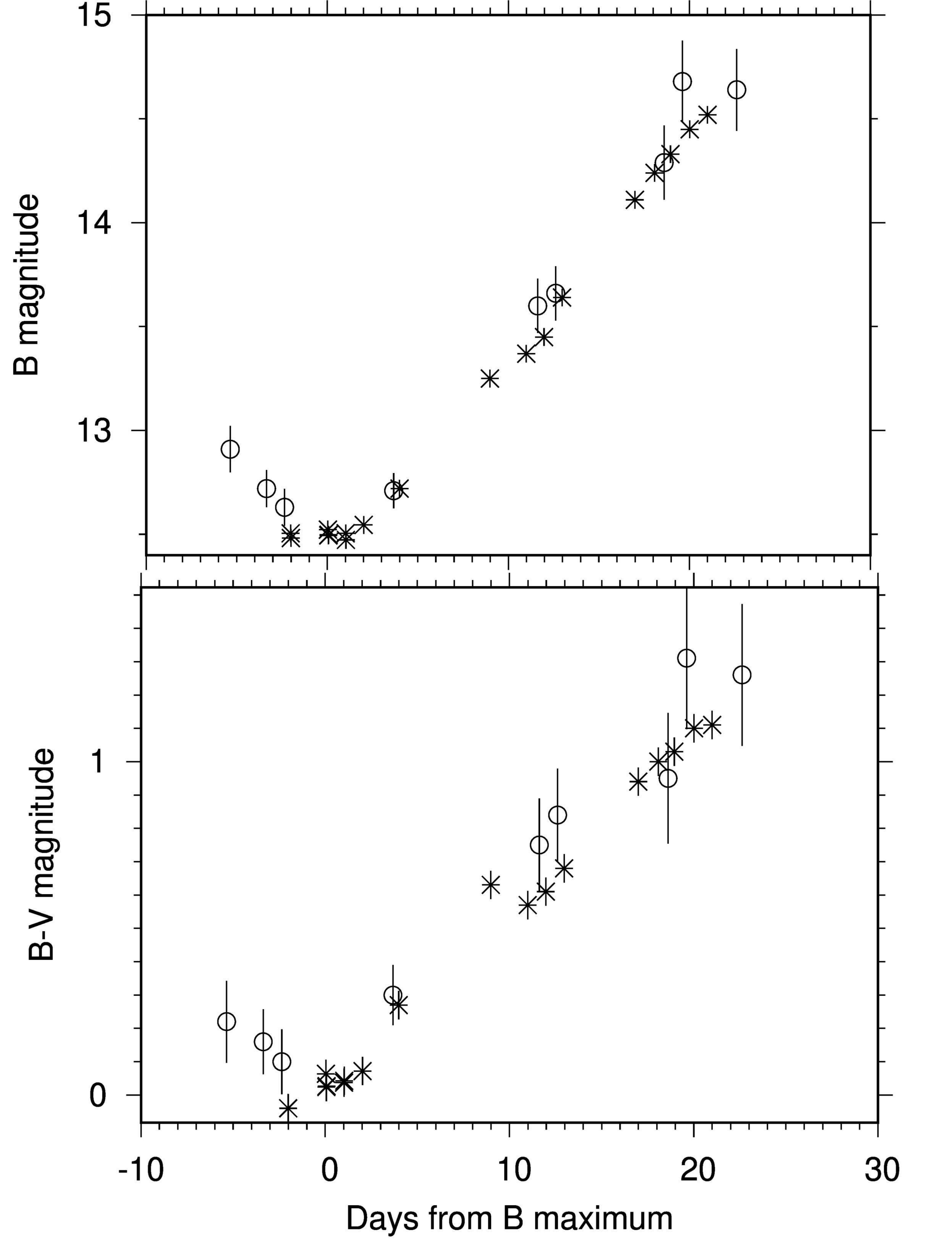}
\vskip -0.3 cm
\caption[]{Light curves of supernovae Ia 1980N and 1981D,
as a function of time since their respective maximum in the B band.
Stars: supernova 1980N.
Open symbols: supernova 1981D.
Top: B magnitudes. Bottom: Difference between magnitudes in the B and V bands.
}
\label{Fig:1980n}
\end{figure}

\subsection*{The 1980N--1981D pair}

Supernovae Ia 1980N and 1981D were both observed in NGC 1316~\cite{Persson:81},
a member of the Fornax cluster. 
As illustrated in Figure~\ref{Fig:1980n}, 
their light curves are remarkably similar~\cite{Persson:81,Hamuy:91}.
Noteworthy, in the B band (Fig.~\ref{Fig:1980n}, top), their 
SALT2 stretches are almost identical,
namely, $x1=-1.14 \pm 0.12$ and $x1=-1.15 \pm 0.36$, respectively,
while the maximum-light magnitudes of both supernovae are the
same within $\pm$~0.1 mag, in the B and V bands~\cite{Hamuy:91} as well as in the J, H, and K ones~\cite{Persson:81}.
Interestingly, 
even though the light curve of 1981D was followed during only 22 days,
the color stretches \cite{Burns:14} of both supernovae look also similar (Fig.~\ref{Fig:1980n}, bottom).

If 1980N and 1981D are two images of the same supernova, according to eqn \ref{eq:dapprox}, given the delay between the two images, the angle between them (see Table~\ref{Table:refsdal}) and the distance to NGC 1316, namely,   
12.5 Mpc \cite{Seibert:21}, the gravitational lens is  
at a distance of 90~$\pm$~1 kpc, that is, 
at the same distance as the outer halo globular cluster
Eridanus, namely, 90.1~kpc \cite{Bellini:19}.

Moreover, according to eqn~\ref{eq:mr}, 
with a mass to radius ratio over 154~$M_\odot R_\odot^{-1}$,
the lens has to be a compact object, like a white dwarf, 
a neutron star or a black hole. 

Typical white dwarfs have mass to radius ratios of
$\approx$~50~$M_\odot R_\odot^{-1}$~\cite{Fontaine:17}.
However, white dwarfs with mass to radius ratios over 300~$M_\odot R_\odot^{-1}$ have been observed~\cite{Fontaine:17,Postnov:20}. 
So, the 1980N--1981D gravitational lens
is likely to be an ultra massive white dwarf.
If so, it is expected to be located almost exactly in the direction where supernova 1981D was observed
($\alpha=$~50.659920$^\circ$, $\delta=-$37.232720$^\circ$).

\begin{figure}[t]
\vskip -0.5 cm
\includegraphics[width=8.0 cm]{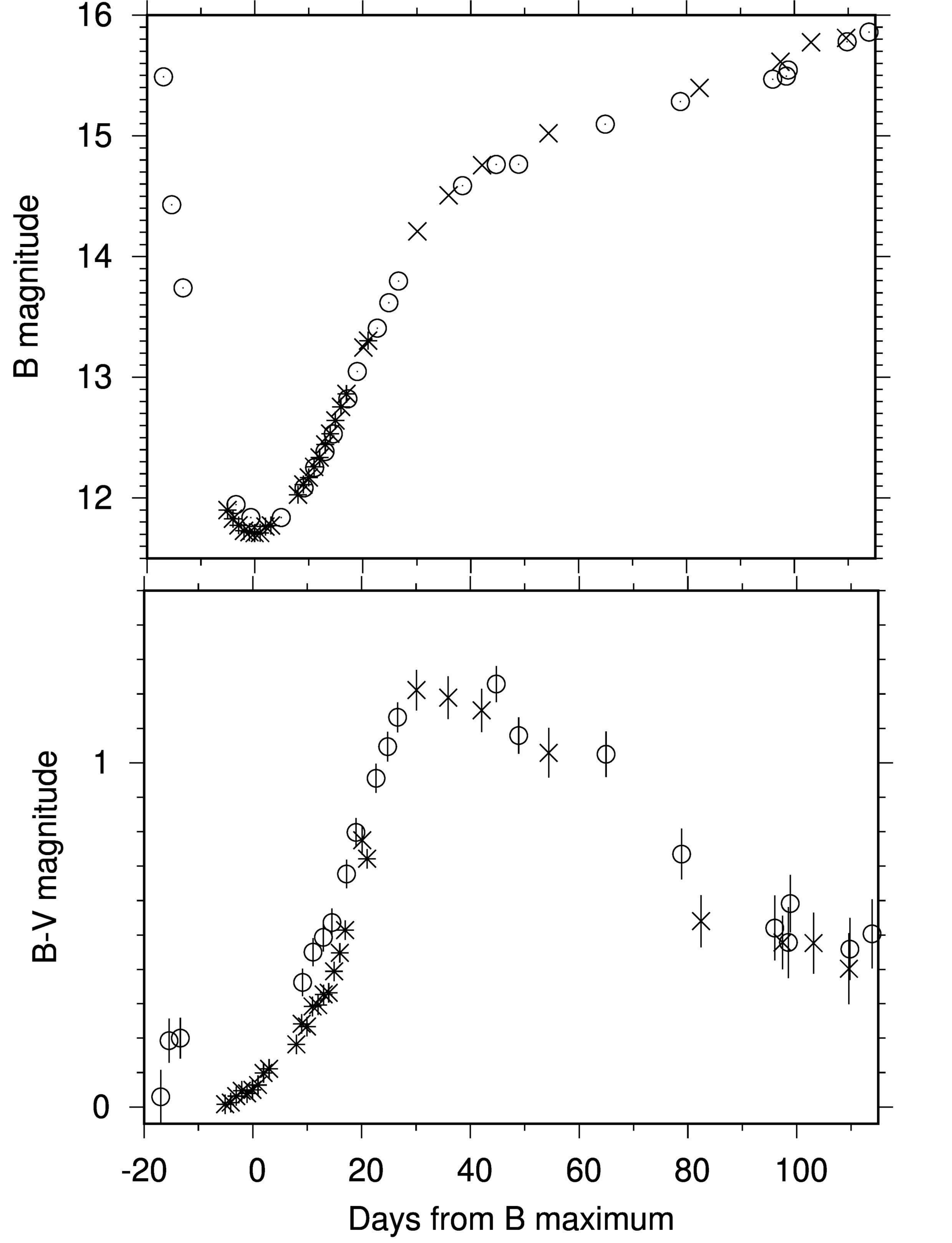}
\vskip -0.3 cm
\caption[]{Light curves of supernovae Ia 2013aa and 2017cbv,
as a function of time since their respective maximum in the B band.
Open symbols: supernova 2017cbv.
For supernova 2013aa, data come from SWIFT (crosses) or from the Carnegie Supernova Project (stars).
Top: B magnitudes (for the sake of clarity, error bars have been omitted). Bottom: Difference between magnitudes in the B and V bands.
}
\label{Fig:2013aa}
\end{figure}

\subsection*{The 2013aa--2017cbv pair}

Supernovae Ia 2013aa and 2017cbv were both observed in NGC 5643 \cite{Vilchez:20}.
As illustrated in Figure~\ref{Fig:2013aa}, 
their light curves are nearly identical \cite{Vilchez:20}.
Noteworthy, in the B band (Fig.~\ref{Fig:2013aa}, top), their 
SALT2 stretches are nearly the same,
namely, $x1=0.63 \pm 0.14$ and $x1=0.62 \pm 0.05$,
respectively\footnote{Data coming from the Carnegie Supernova Project, during which both supernovae were studied.}. 
Interestingly, 
their color stretches seem also close (Fig.~\ref{Fig:2013aa}, bottom).
Moreover, their spectra are remarkably similar \cite{Vilchez:20}.

If 2013aa and 2017cbv are two images of the same supernova, according to eqn \ref{eq:dapprox}, given the delay between the two images, the angle between them (Table \ref{Table:refsdal}) and the distance to NGC 5643, namely,   
18.8 Mpc \cite{Seibert:21}, the gravitational lens is  
at a distance of 702~$\pm$ 1 kpc, that is, 
nearly at the same distance as the Andromeda galaxy \cite{Guinan:10}.

Moreover, according to eqn~\ref{eq:mr}, 
with a mass to radius ratio over 193~$M_\odot R_\odot^{-1}$,
the 2013aa--2017cbv gravitational lens is also likely to be an ultra massive white dwarf. 
If so, it is expected to be located almost exactly in the direction where supernova 2017cbv was observed
($\alpha=$~218.143420$^\circ$, $\delta=-$44.134090$^\circ$).

\section*{Discussion}
 
\subsection*{Where are both lenses ?}

Lens 1980N--1981D is roughly in the same direction
as the Fornax dwarf spheroidal galaxy 
($\alpha=~$39.997200$^\circ$, $\delta=-$34.449187$^\circ$).
However the later is significantly farer,
namely, at a distance of 143$~\pm~$3 kpc \cite{Seibert:22}.
On the other hand, 
it is at the same distance as the outer
halo globular cluster Eridanus 
($\alpha=$~66.185417$^\circ$, $\delta=-$21.186943$^\circ$) \cite{Bellini:19}.
However, it is more than 30 kpc away,
far from and not in line with its tidal tails \cite{Dacosta:17}.

Lens 2013aa--2017cbv seems even farer from any known galaxy of the Local Group but, given its large distance, 
it could for instance belong to some yet unknown cluster or stream of stars.

\subsection*{Are both lenses observable ?}

Both lenses are quite far, namely, 
at a distance of $\approx$ 90 and 700 kpc, respectively
(Table \ref{Table:refsdal}),
while most studies 
of remote white dwarfs 
have focused on cases located in rather close 
globular or open star clusters \cite{Campos:16,Kilic:18,Thiele:21}.
However, an accreting white dwarf 
has already been detected in the Andromeda galaxy,
thanks to its recurrent X-ray emission \cite{Tang:14}.
Note also that LISA has the potential to detect detached double white dwarf binaries in neighboring galaxies, up to the border of the Local Group \cite{Rossi:18}.

\subsection*{How many of them ?}

In the present study, pairs of images of the same supernova have been looked for using rather strict criteria, meaning that some pairs may have been overlooked.
For instance, comparison of host masses was considered as a selection criterion,
but host masses have not been recorded
for all supernovae of the Pantheon+ sample. 
Moreover, it has been assumed that errors on host mass estimates can be safely trusted. 
In this respect, note that
it would be more rigorous to compare the total surface brightnesses of both hosts, since 
this is the quantity that is expected to be conserved 
upon light deflection \cite{Virbhadra:22}.

\section*{Conclusion}

Two likely pairs of images have been found in the Pantheon+ sample of supernovae Ia, that is, two pairs of supernovae with strikingly similar light curves \cite{Hamuy:91,Vilchez:20} (Fig. \ref{Fig:1980n} and \ref{Fig:2013aa}). In both cases, the gravitational lens has to be a compact object with a mass to radius ratio over 150~$M_\odot R_\odot^{-1}$ (Table \ref{Table:refsdal}), being likely an ultra massive white dwarf. While lens 1980N--1981D is located in the outer halo of the Milky Way, lens 2013aa--2017cbv is at the same distance as the Andromeda galaxy.  

Masses of nearby white dwarfs 
were previously measured using related effects, like
astrometric microlensing \cite{Livio:17} or self-lensing \cite{Wako:19}.
However, to my knowledge, this is the first
time strong gravitational lensing is used for measuring
the distance and estimating the mass to radius ratio of a star.

Using this approach,
ongoing surveys of supernovae Ia, like the Zwicky Transient Facility \cite{Dhawan:22} or the Dark Energy Survey \cite{Abbott:24}, are expected to provide more cases. 
Of course, compact lenses in the Local Group could also be found 
by studying other kinds of well-characterized transients.

With more statistics,
the study of the population 
of compact objects in the halo of the Milky Way or within the Local Group
may prove interesting, for instance for gaining 
better insights about the substructures of the former or the history of the later. Understanding the origin of apparently isolated objects like those discovered in the course of the present study could also reveal key features of the dynamics of the Local Group.

\vskip 0.5cm
\noindent


\begin{thebibliography}{10}
\providecommand{\url}[1]{\texttt{#1}}
\providecommand{\urlprefix}{}
\providecommand{\eprint}[2][]{\url{#2}}

\bibitem{Weymann:79}
Walsh, D., Carswell, R.F. \& Weymann, R.J. (1979).
\newblock {0957+561 A, B: twin quasistellar objects or gravitational lens ?}
\newblock \emph{Nature} \textbf{279}(5712), 381--384.

\bibitem{Schramm:96}
Pelt, J., Kayser, R., Refsdal, S. \& Schramm, T. (1996).
\newblock {The light curve and the time delay of QSO 0957+561.}
\newblock \emph{Astron. Astrophys.} \textbf{305}, 97.

\bibitem{Oscoz:01}
Oscoz, A., Alcalde, D., Serra-Ricart, M., Mediavilla, E., Abajas, C., Barrena,
  R., Licandro, J., Motta, V. \& Munoz, J. (2001).
\newblock {Time delay in QSO 0957+561 from 1984--1999 optical data}.
\newblock \emph{Ap. J.} \textbf{552}(1), 81.

\bibitem{Refsdal:64}
Refsdal, S. (1964).
\newblock {On the possibility of determining Hubble's parameter and the masses
  of galaxies from the gravitational lens effect}.
\newblock \emph{Mon. Not. R. Astron. Soc.} \textbf{128}(4), 307--310.

\bibitem{Auger:09}
Dobke, B.M., King, L.J., Fassnacht, C.D. \& Auger, M.W. (2009).
\newblock Estimating cosmological parameters from future gravitational lens
  surveys.
\newblock \emph{Mon. Not. R. Astron. Soc.} \textbf{397}(1), 311--319.

\bibitem{Meylan:20}
Wong, K.C., Suyu, S.H., Chen, G.C., Rusu, C.E., Millon, M., Sluse, D., Bonvin,
  V., Fassnacht, C.D., Taubenberger, S., Auger, M.W. \emph{et~al.} (2020).
\newblock {H0LiCOW XIII. A 2.4\% measurement of H0 from lensed quasars:
  5.3$\sigma$ tension between early and late-Universe probes}.
\newblock \emph{Mon. Not. R. Astron. Soc.} \textbf{498}, 1420--1439.

\bibitem{Riess:22}
Riess, A.G., Yuan, W., Macri, L.M., Scolnic, D., Brout, D., Casertano, S.,
  Jones, D.O., Murakami, Y., Anand, G.S., Breuval, L. \emph{et~al.} (2022).
\newblock {A comprehensive measurement of the local value of the Hubble
  constant with 1 km s$^{-1}$ Mpc$^{-1}$ uncertainty from the Hubble Space
  Telescope and the SH0ES team}.
\newblock \emph{Ap. J. letters} \textbf{934}(1), L7.

\bibitem{Tucker:15}
Kelly, P.L., Rodney, S.A., Treu, T., Foley, R.J., Brammer, G., Schmidt, K.B.,
  Zitrin, A., Sonnenfeld, A., Strolger, L.G., Graur, O. \emph{et~al.} (2015).
\newblock Multiple images of a highly magnified supernova formed by an
  early-type cluster galaxy lens.
\newblock \emph{Science} \textbf{347}(6226), 1123--1126.

\bibitem{Zitrin:16}
Kelly, P.L., Rodney, S., Treu, T., Strolger, L.G., Foley, R., Jha, S., Selsing,
  J., Brammer, G., Brada{\v{c}}, M., Cenko, S.B. \emph{et~al.} (2016).
\newblock Deja vu all over again: the reappearance of supernova refsdal.
\newblock \emph{Ap. J. letters} \textbf{819}(1), L8.

\bibitem{Whitaker:21}
Rodney, S.A., Brammer, G.B., Pierel, J.D., Richard, J., Toft, S., O'Connor,
  K.F., Akhshik, M. \& Whitaker, K.E. (2021).
\newblock A gravitationally lensed supernova with an observable two-decade time
  delay.
\newblock \emph{Nature Astronomy} \textbf{5}(11), 1118--1125.

\bibitem{Scolnic:22}
Scolnic, D., Brout, D., Carr, A., Riess, A.G., Davis, T.M., Dwomoh, A., Jones,
  D.O., Ali, N., Charvu, P., Chen, R. \emph{et~al.} (2022).
\newblock {The Pantheon+ analysis: the full data set and light-curve release}.
\newblock \emph{Ap. J.} \textbf{938}(2), 113.

\bibitem{Sanejouand:23}
Sanejouand, Y.H. (2023).
\newblock {A robust assessment of the local anisotropy of the Hubble constant}.
\newblock \emph{arXiv} \textbf{2312}, 05896.

\bibitem{Scolnic:20}
Scolnic, D., Smith, M., Massiah, A., Wiseman, P., Brout, D., Kessler, R.,
  Davis, T., Foley, R., Galbany, L., Hinton, S.R. \emph{et~al.} (2020).
\newblock {Supernova siblings: assessing the consistency of properties of type
  Ia supernovae that share the same parent galaxies}.
\newblock \emph{Ap. J. letters} \textbf{896}(1), L13.

\bibitem{Persson:81}
Elias, J.H., Frogel, J.A., Hackwell, J.A. \& Persson, S.E. (1981).
\newblock {Infrared light curves of Type I supernovae}.
\newblock \emph{Astrophys. J.} \textbf{251}, L13--L16.

\bibitem{Hamuy:91}
Hamuy, M., Phillips, M., Maza, J., Wischnjewsky, M., Uomoto, A., Landolt, A.U.
  \& Khatwani, R. (1991).
\newblock {The optical light curves of SN 1980N and SN 1981D in NGC 1316
  (Fornax A)}.
\newblock \emph{A. J.} \textbf{102}, 208--217.

\bibitem{Burns:14}
Burns, C.R., Stritzinger, M., Phillips, M., Hsiao, E., Contreras, C., Persson,
  S., Folatelli, G., Boldt, L., Campillay, A., Castell{\'o}n, S. \emph{et~al.}
  (2014).
\newblock {The Carnegie supernova project: intrinsic colors of type Ia
  supernovae}.
\newblock \emph{Ap. J.} \textbf{789}(1), 32.

\bibitem{Seibert:21}
Hoyt, T.J., Beaton, R.L., Freedman, W.L., Jang, I.S., Lee, M.G., Madore, B.F.,
  Monson, A.J., Neeley, J.R., Rich, J.A. \& Seibert, M. (2021).
\newblock {The carnegie chicago hubble program X: Tip of the red giant branch
  distances to NGC 5643 and NGC 1404}.
\newblock \emph{Ap. J.} \textbf{915}(1), 34.

\bibitem{Bellini:19}
Baumgardt, H., Hilker, M., Sollima, A. \& Bellini, A. (2019).
\newblock {Mean proper motions, space orbits, and velocity dispersion profiles
  of Galactic globular clusters derived from Gaia DR2 data}.
\newblock \emph{Mon. Not. R. Astron. Soc.} \textbf{482}(4), 5138--5155.

\bibitem{Fontaine:17}
B{\'e}dard, A., Bergeron, P. \& Fontaine, G. (2017).
\newblock Measurements of physical parameters of white dwarfs: A test of the
  mass--radius relation.
\newblock \emph{Ap. J.} \textbf{848}(1), 11.

\bibitem{Postnov:20}
Pshirkov, M.S., Dodin, A.V., Belinski, A.A., Zheltoukhov, S.G., Fedoteva, A.A.,
  Voziakova, O.V., Potanin, S.A., Blinnikov, S.I. \& Postnov, K.A. (2020).
\newblock {Discovery of a hot ultramassive rapidly rotating DBA white dwarf}.
\newblock \emph{Mon. Not. R. Astron. Soc. lett.} \textbf{499}(1), L21--L25.

\bibitem{Vilchez:20}
Burns, C.R., Ashall, C., Contreras, C., Brown, P., Stritzinger, M., Phillips,
  M., Flores, R., Suntzeff, N.B., Hsiao, E.Y., Uddin, S. \emph{et~al.} (2020).
\newblock {SN 2013aa and SN 2017cbv: Two Sibling Type Ia Supernovae in the
  spiral galaxy NGC 5643}.
\newblock \emph{Ap. J.} \textbf{895}(2), 118.

\bibitem{Guinan:10}
Vilardell, F., Ribas, I., Jordi, C., Fitzpatrick, E.L. \& Guinan, E.F. (2010).
\newblock {The distance to the Andromeda galaxy from eclipsing binaries}.
\newblock \emph{Astronomy \& Astrophysics} \textbf{509}, A70.

\bibitem{Seibert:22}
Oakes, E.K., Hoyt, T.J., Freedman, W.L., Madore, B.F., Tran, Q.H., Cerny, W.,
  Beaton, R.L. \& Seibert, M. (2022).
\newblock {Distances to Local Group Galaxies via Population II, Stellar
  Distance Indicators. II. The Fornax Dwarf Spheroidal}.
\newblock \emph{Ap. J.} \textbf{929}(2), 116.

\bibitem{Dacosta:17}
Myeong, G., Jerjen, H., Mackey, D. \& Da~Costa, G.S. (2017).
\newblock {Tidal Tails around the Outer Halo Globular Clusters Eridanus and
  Palomar 15}.
\newblock \emph{Ap. J. letters} \textbf{840}(2), L25.

\bibitem{Campos:16}
Campos, F., Bergeron, P., Romero, A.D., Kepler, S.O., Ourique, G., Costa,
  J.E.d.S., Bonatto, C.J., Winget, D.E., Montgomery, M.H., Pacheco, T.A.
  \emph{et~al.} (2016).
\newblock A comparative analysis of the observed white dwarf cooling sequence
  from globular clusters.
\newblock \emph{Mon. Not. R. Astron. Soc.} \textbf{456}(4), 3729--3742.

\bibitem{Kilic:18}
Williams, K.A., Canton, P.A., Bellini, A., Bolte, M., Rubin, K.H., Gianninas,
  A. \& Kilic, M. (2018).
\newblock {Ensemble properties of the white dwarf population of the old, solar
  metallicity open star cluster Messier 67}.
\newblock \emph{Ap. J.} \textbf{867}(1), 62.

\bibitem{Thiele:21}
Richer, H.B., Caiazzo, I., Du, H., Grondin, S., Hegarty, J., Heyl, J., Kerr,
  R., Miller, D.R. \& Thiele, S. (2021).
\newblock Massive white dwarfs in young star clusters.
\newblock \emph{Ap. J.} \textbf{912}(2), 165.

\bibitem{Tang:14}
Tang, S., Bildsten, L., Wolf, W.M., Li, K., Kong, A.K., Cao, Y., Cenko, S.B.,
  De~Cia, A., Kasliwal, M.M., Kulkarni, S.R. \emph{et~al.} (2014).
\newblock {An accreting white dwarf near the Chandrasekhar limit in the
  Andromeda galaxy}.
\newblock \emph{Ap. J.} \textbf{786}(1), 61.

\bibitem{Rossi:18}
Korol, V., Koop, O. \& Rossi, E.M. (2018).
\newblock {Detectability of double white dwarfs in the local group with LISA}.
\newblock \emph{Ap. J. letters} \textbf{866}(2), L20.

\bibitem{Virbhadra:22}
Virbhadra, K. (2022).
\newblock {Distortions of images of Schwarzschild lensing}.
\newblock \emph{Phys. Rev. D} \textbf{106}(6), 064038.

\bibitem{Livio:17}
Sahu, K.C., Anderson, J., Casertano, S., Bond, H.E., Bergeron, P., Nelan, E.P.,
  Pueyo, L., Brown, T.M., Bellini, A., Levay, Z.G. \emph{et~al.} (2017).
\newblock Relativistic deflection of background starlight measures the mass of
  a nearby white dwarf star.
\newblock \emph{Science} \textbf{356}(6342), 1046--1050.

\bibitem{Wako:19}
Masuda, K., Kawahara, H., Latham, D.W., Bieryla, A., Kunitomo, M., MacLeod, M.
  \& Aoki, W. (2019).
\newblock {Self-lensing discovery of a 0.2 M$_\odot$ white dwarf in an
  unusually wide orbit around a Sun-like star}.
\newblock \emph{Ap. J.} \textbf{881}(1), L3.

\bibitem{Dhawan:22}
Dhawan, S., Goobar, A., Smith, M., Johansson, J., Rigault, M., Nordin, J.,
  Biswas, R., Goldstein, D., Nugent, P., Kim, Y. \emph{et~al.} (2022).
\newblock {The Zwicky Transient Facility Type Ia supernova survey: first data
  release and results}.
\newblock \emph{Mon. Not. R. Astron. Soc.} \textbf{510}(2), 2228--2241.

\bibitem{Abbott:24}
Abbott, T.M.C., Acevedo, M., Aguena, M., Alarcon, A., Allam, S., Alves, O.,
  Amon, A., Andrade-Oliveira, F., Annis, J., Armstrong, P. \emph{et~al.}
  (2024).
\newblock {The Dark Energy Survey: Cosmology Results With $\approx$ 1500 New
  High-redshift Type Ia Supernovae Using The Full 5-year Dataset}.
\newblock \emph{arXiv} \textbf{2401}, 02929.

\end{thebibliography}

\end{document}